\documentclass[conference]{IEEEtran}

\usepackage[english]{babel}
\usepackage{blindtext}
\usepackage{listings}
\usepackage{xcolor}
\usepackage{paralist}
\usepackage{booktabs}
\usepackage{courier}
\usepackage{tabularx}
\usepackage{makecell}
\usepackage{enumitem}
\usepackage{amsfonts}
\usepackage{amssymb}
\usepackage{graphicx}

\newcommand{\bbR}{\mathbb{R}}

\begin{document}

\title{The Scalability-Efficiency/Maintainability-Portability Trade-off in Simulation Software Engineering: \\ 
Examples and a Preliminary Systematic Literature Review}

\author{\IEEEauthorblockN{Dirk Pfl{\"u}ger, Miriam Mehl,\\Julian Valentin, Florian Lindner,\\David Pfander}
\IEEEauthorblockA{Institute for Parallel and Distributed Systems\\
University of Stuttgart, Germany\\
Email: \{dirk.pflueger, miriam.mehl,\\julian.valentin, florian.lindner,\\ david.pfander\}@ipvs.uni-stuttgart.de}
\and
\IEEEauthorblockN{Stefan Wagner, Daniel Graziotin,\\Yang Wang}
\IEEEauthorblockA{Institute of Software Technology\\
University of Stuttgart, Germany\\
Email: \{stefan.wagner, daniel.graziotin,\\yang.wang\}@iste.uni-stuttgart.de}
}

\maketitle

\begin{abstract}
Large-scale simulations play a central role in science and the industry. 
Several challenges occur when building simulation software, because simulations require complex software developed in a dynamic construction process. That is why simulation software engineering (SSE) is emerging lately as a research focus. The dichotomous trade-off between scalability and efficiency (SE) on the one hand and maintainability and portability (MP) on the other hand is one of the core challenges. We report on the SE/MP trade-off in the context of an ongoing systematic literature review (SLR). After characterizing the issue of the SE/MP trade-off using two examples from our own research, we (1) review the 33 identified articles that assess the trade-off, (2) summarize the proposed solutions for the trade-off, and (3) discuss the findings for SSE and future work. 
Overall, we see evidence for the SE/MP trade-off and first solution approaches. However, a strong empirical foundation has yet to be 
established; general quantitative metrics and methods supporting software developers in addressing the trade-off have to be developed. 
We foresee considerable future work in SSE across scientific communities.
\end{abstract}

\IEEEpeerreviewmaketitle




\section{Introduction}

We consider simulation software engineering (SSE) from a two-fold perspective: the perspective of software developers specialized in 
methods of scientific computing and high performance computing and the perspective of software engineering. 

\subsection{Problem Statement}

Our experiences as scientific software developers show that software engineering tools and methods are used and are very helpful for various tasks 
such as testing, version control, debugging, and the design of the object-oriented software architecture. However, a question that 
is of particular importance for high performance simulation software is not addressed by the available software engineering research: 
How do we achieve a high hardware efficiency and parallel scalability on massively parallel and heterogeneous supercomputers without 
deteriorating the readability, maintainability, and portability of the code? Whereas efficiency and scalability often require 
a tailoring of the code to specific hardware details such as cache line lengths and memory bandwidth and, therefore, the use of 
low-level languages, maintainability and portability require a certain abstraction level from the actual hardware and, thus, e.g., 
a modular software architecture, clear separation of concerns, and the restriction to standard language elements. 

Moreover, simulation software is different from many other software types from the software engineering perspective: The 
aim of the software is often ill-defined in the early phases, there is no clear separation of roles between programmers and customers, fundamental 
parts of the code have to be reimplemented frequently due to new scientific and methodological insights, the requirements of scientific
projects to produce results quickly do not allow for extended phases of software development before reaching a first usable software prototype,
a missing or fuzzy knowledge of the correct behavior of the software makes testing and verification much harder, {\em and} the fact that
there actually is a trade-off and not a synergy between scalability/efficiency and maintainability/portability is different from, e.g.,
software for embedded systems, where a high abstraction can help to reach a high efficiency. 

\subsection{Research Objective}

Whereas other literature reviews such as, in particular, \cite{farhoodi13}, cover the total view on software engineering for 
scientific software, we narrow the focus on 1) a special type of scientific software, i.e., simulation software intended to run 
efficiently on massively parallel high performance architectures, 2) the trade-off between scalability \& efficiency on the one
hand and maintainability \& portability on the other hand. This SE/MP trade-off seems to be one of the most central issues in
software development in this area with only few systematic solution approaches available. As an initial step for research on this 
specific topic, we report exemplary experiences from our own research on simulation software and, as the main part of the
paper, present results and conclusions from a systematic literature review to get a clear view of the state-of-the-art.

\subsection{Definitions}
In literature, many different terms are used to describe research fields and topics related to 
this paper. We give our definitions here which we adhere to for the rest of the paper.\\
\textbf{High Performance Computing:} High performance computing (HPC) is the area of 
computer-supported simulation on massively parallel computing architectures with particular high requirements in terms of computational performance and memory efficiency.\\
\textbf{Simulation Software:} This type of software aims at modeling and analyzing scientific problems, e.g., from 
engineering, physics, chemical science, biology, or medicine.\\ 
\textbf{Simulation Software Engineering:} In SSE, we focus on
all activities that are concerned with developing a simulation software.

\section{Experiences}

Experiences from the research projects of the authors \cite{P18} show that the trade-off between scalability and efficiency on the 
one hand and maintainability and portability on the other hand is ubiquitous in scientific simulation software development. In this section, we
present two examples that represent two extremes in terms of granularity: a software environment for complex
multi-physics simulations and code snippets for matrix-matrix multiplication, one of the low-level building blocks of many simulation codes.

\subsection{Multiphysics Simulations}

Multi-physics simulations are characterized by the requirement to combine physical phenomena described by different mathematical models in a single simulation. With increasing compute power, improved numerical solvers and increasing model accuracy requirements, they have become a very important class of scientific simulations during the last decade. Two main approaches to 
develop corresponding simulation software have been established:
monolithic approaches that tackle all equations in a single system and, thus, a single software, and partitioned approaches that re-use existing software for the 
involved single-physics sub-problems. Simple combinatorial consideration shows that already for five sub-problems, ten different 
monolithic codes for pairs of them and $31$ different codes for all possible combinations would have to be developed from scratch. Thus, 
a partitioned, i.e., highly modular approach seems to be very attractive. The partitioned approach, however, raises questions on whether the same numerical and hardware efficiency and parallel scalability as for monolithic codes can be achieved. 

The software library preCICE \cite{precice} has been developed (in a small group of doctoral students with a varying number of members between one and three) as a tool for surface coupling between sub-domains with different physics. It is written in standard C++ using MPI parallelization and implements point-to-point communication between parallel single-physics solvers, data mapping between non-matching meshes and outer iterative solvers recovering the monolithic solution for implicit time-stepping or steady-state scenarios. Our standard application example is the interaction of fluid flow with elastic structures. 

In the following, we summarize results from previous publications and interpret them with the focus on the trade-off between scalability/efficiency and maintainability/portability. 

{\bf Scalability \& Efficiency.} 
The major concern in partitioned multi-physics simulation approaches is that they might degrade the scalability and efficiency both in terms of numerical efficiency and in terms of hardware performance. To show that this is not necessarily true, we simulated 1) a
Gaussian pulse moving in an artificially split domain and 2) various real fluid-structure interactions. 

The simulations showed that,
given an efficient realization of inter-code communication, scalability and efficiency losses due to the domain splitting are low. The coupled simulation inherits the high scalability and efficiency of the coupled simulation codes \cite{precice}.\footnote{The simulation was done with the high order discontinuous Galerkin code Ateles on octree meshes \cite{Apes}.} With 2), we can prove that only a very moderate number of outer iterations is required for real-world problems. This results in an increase of computational cost that is similar to what we observe in monolithic approaches due to the worse condition of the coupled system compared to the condition of the single-physics problems \cite{FSI}.

{\bf Maintainability \& Portability.} The high number of commercial and academic codes (2 commercial, 13 academic codes, \cite{precice}) that have been coupled with preCICE in the last 5 years based on the work of only a handful of doctoral students hints at an enormous reduction of development time compared to writing a new code from scratch. In addition, the resulting flexibility in exchanging coupled codes allows us to use single-physics solvers optimized for the respective target architectures. The preCICE code itself is highly portable as it uses only standard C++ and MPI functionality and has only a low number of dependencies 
\cite{precice}. It is strictly object-oriented and of moderate size. Experiences with modifications or integration of new methods over the last five years showed that due to this not only the maintainability of the whole coupled simulation environment but also of preCICE itself is very good. 

{\bf Conclusion.} Other than in the example presented in \ref{subsec:matvec}, a highly modular approach at the level of multiple interacting physical effects in a simulation scenario seems to be very appropriate in terms of maintainability \& portability without degenerating scalability \& efficiency.

\subsection{Matrix-Matrix Multiplication}
\label{subsec:matvec}

Matrix-matrix and matrix-vector operations are the core operations in
simulation codes. The underlying physics is often modeled
as a system of partial or ordinary differential equations. Their discretization via finite elements or finite differences often
leads to a (sparse) system of linear equations. 
If an iterative solver is employed, 
the expensive core operation typically is a matrix-vector multiplication. 
In addition, matrix-matrix multiplications occur, e.g., when transformations
are applied to the discretized system.

In the following, we consider the multiplication of two dense square matrices. 
From a programmer's point of view, this should not be very challenging, easy to implement,
maintainable and portable, and well supported by modern hardware and
compilers. As we will show, this is true with respect to the ease of
programming, maintainability \& portability. However, it turns out that the easy implementation leads to very poor efficiency. The more a
programmer reaches towards the hardware's peak performance, the more
specialized and the less maintainable and portable the code gets. This
is what we demonstrate in the following. As the target hardware, we
consider an Intel i5 4300 Laptop CPU of the Mobile Processor
Series. It exhibits two cores, 2.6\,GHz clock cycle, 2 vector units,
support for fused-multiply-add operations and 4-wide vector
units yielding a theoretical peak performance (TPP) of 83.2\,GFlops (double
precision).

The computation of 
\[
 A B = C, \quad A,B,C \in \bbR^{n\times n},
\]
with $n=4096$ for the following performance results, can easily be
implemented in four lines of code, see Listing~\ref{lst:naive}. This is clearly maintainable \& portable. However,
it results in a performance of a mere 0.22\,GFlops, which
corresponds to the TPP of an Intel Pentium 200 from 1996.

\begin{lstlisting}[label=lst:naive,float=htpb,
caption={Direct implementation}]  % Start your code-block

for (size_t i = 0; i < n; i += 1) {
 for (size_t j = 0; j < n; j += 1) {
  for (size_t k = 0; k < n; k += 1) {
   C[i * n + j] += A[i * n + k] * B[k * n + j]; 
  }
 }
}
\end{lstlisting}

This is a classical example in HPC lectures: The memory access to
matrix $B$ is non-contiguous. This hinders prefetching and
leads to a significant pollution of the hardware caches. A simple
transposed access of B comes to the rescue, which compilers do
not do automatically. This increases the performance to 0.5\,GFlops
(corresponding roughly to the TPP of a Pentium2 450 from 1998). However, we have
then tailored our implementation to the memory layout of the
programming language $C/C++$.

Compilers cannot detect that the function can be vectorized. We can
either enforce auto-vectorization by replacing the dependency on the matrix $C$ with
an accumulator variable which leads to 1.04\,GFlops, or we 
can directly use the shared memory parallelization of OpenMP, which
can then make use of both cores and two processes per core
(Hyper-Threading). A simple pragma statement is sufficient, see
Listing~\ref{lst:omp_parfor}, which leads to 3.23\,GFlops (the
TPP of a Pentium 4 from 2001).

\begin{lstlisting}[label=lst:omp_parfor,float=htpb, %frame=single,
caption={Shared-memory parallelization with OpenMP}]  % Start your code-block

#pragma omp parallel for
for (size_t i = 0; i < n; i++) {
  // ...
}
\end{lstlisting}

Analyzing the suboptimal performance, we realize that memory access is
the main bottleneck. The main memory can be accessed with a maximum of
25.6\,GB/s on our i5 4300 chipset. We can thus only read about 0.32 Bytes
per floating point operation. This is not sufficient to provide enough
data in time. We therefore have to rewrite our code so that we can
execute as many computations as possible for each loaded batch of
data. The idea is to work on sub-blocks (sub-matrices) that fit into the
cache of our system. As a side-effect, we obtain a performance that is
independent of the matrix size $n$, but we loose auto-vectorization
once more. The resulting code is now 33 lines of code and much less
maintainable (see Listing~\ref{lst:blocking} for some hints). But our
performance jumps to 12.7\,GFlops when we combine blocking and OpenMP. This
corresponds to the TPP of an Intel Core 2 Duo E6600 from 2006.

\begin{lstlisting}[label=lst:blocking,float=htpb,
caption={A blocked implementation using OpenMP}]  % Start your code-block

#define BLOCKSIZE 128
#define KCHUNK 128
omp_set_num_threads(4);

#pragma omp parallel for
for (size_t i = 0; i < n; i += BLOCKSIZE) {
 for (size_t j = 0; j < n; j += BLOCKSIZE) {
  double result[BLOCKSIZE * BLOCKSIZE];
  // ...
  for (size_t kBlock=0;kBlock<n;kBlock+=KCHUNK) {
   // ...
   // Inner block multiplication
   for (size_t ii = 0; ii < BLOCKSIZE; ii += 1) {
    for (size_t jj = 0; jj< BLOCKSIZE; jj += 1) {
     double resultChunk1[4]; // only one register
     // ...
     for (size_t k = 0; k < KCHUNK; k += 4) {
      for (size_t kk = 0; kk < 4; kk++) { 
       resultChunk1[kk] += AA[(ii+0)*KCHUNK+k+kk] 
                         * BB[(jj+0)*KCHUNK+k+kk];
      }
     }
     // ...
\end{lstlisting}

Remembering that we do not exploit the vectorization that our
processor provides, we manually vectorize our code using
intrinsics. This results in 49 lines of code of which we show three lines in
Listing~\ref{lst:intrinsics}. The
resulting implementation is not intuitively understandable
any more. However, we now reach 38\,GFlops in total, thus about the
TPP of an Intel Core 2 Quad Q6700 from
2007.

{\bf Conclusion.} Reaching to almost 50\% of the theoretical peak performance on
our system is an excellent result for the performance
optimization. But the resulting code is custom-tailored to a certain
hardware platform, exploits specific properties such as the width of
vector units and the size of cache-lines, and can result in
significant performance penalties on other hardware -- if it
compiles at all. This example demonstrates that we need new ways to
obtain an acceptable trade-off between computational efficiency 
on the one hand and maintainability and portability on the other hand~\cite{heinecke13emerging,heinecke13demonstrating}.

\begin{lstlisting}[label=lst:intrinsics,float=htpb,
caption={Three lines of the inner-most loop of an implementation with OpenMP and Intrinsics}]  % Start your code-block

 // ...
 alignas(32) __m256d Aregister[REG_BLOCK_II];
 for (size_t iii=0; iii<REG_BLOCK_II; iii++) {
   Aregister[iii] = _mm256_set1_pd(
        AA[(ii + iii) * KCHUNK + k + 0]);
 // ...
\end{lstlisting}

\section{Review Questions}
To put our own experiences described above into a broader perspective, we performed a
SLR. We started with a broader set of review questions related to understanding the emerging 
field of SSE as defined in \cite{P18},
as well as the corresponding challenges and advantages. For the present paper, we focused only on
the SE/MP trade-off and formulated the following questions:

\begin{compactitem}
\item What are issues around scalability, efficiency, maintainability and portability in SSE?
\item How is the SE/MP trade-off addressed?
\item How valid/quantitative is the knowledge we have on the SE/MP trade-off?
\item Which methods and solutions are proposed?
\end{compactitem}

\vspace{2em}

\section{Method}\label{method-1-page}

We followed the established guidelines for SLRs
proposed by Kitchenham \cite{Kitchenham:2007}.

\subsection{Data Sources and Search Strategy\label{ssec:data-sources}}

We searched the papers using a three-fold strategy for data sources. 
\begin{enumerate}[noitemsep, nosep]
\item \label{itm:search1} Academic search engines and digital libraries (IEEE Xplore, ACM DL, ScienceDirect, Web of Science, SEI Digital Library, Wiley Interscience, and Inspec). 
\item \label{itm:search2}Manual inspection of the output of the publication venues that are known to host potentially relevant articles (the proceedings of the ``Software Engineering for Computational Science Workshop'' series, the ``Computing in Science and Engineering'',  ``International Journal of High Performance Computing Applications'', and ``IEEE Software'' journals).
\item \label{itm:search3} Manual inspection of the results in steps \ref{itm:search1} and \ref{itm:search2}, after applying inclusion and exclusion criteria (see \ref{ssec:inclusion-exclusion}, \ref{ssec:se-mp-trade-off} for articles that reflect about the SE/MP trade-off).
\end{enumerate}

We constructed the search query following the quasi-gold standard (QGS) strategy suggested by Zhang et al.~\cite{Zhang:2011ff}. The  strategy aims to raise the objectivity of the search queries by extracting the search terms from a collection of articles that are considered as the gold standard of a given topic. The quasi part of QGS is related to the time and context dependency that are typical in software engineering. The constructed queries, when run, should attempt to discover as much of the QGS collection as possible. We defined a QGS from 69 articles, which let us construct the following query. 

\textit{( ``software engineering'' OR  ``software development'') AND ( ``simulation software'' OR  ``simulation code''`) AND ( ``scientific computing'' OR  ``scientific software'' OR  ``computational science'' OR  ``computational simulation'' OR  ``high performance computing'' OR  ``parallel computing'')}

One author of the present paper performed the search for the articles.

\subsection{Inclusion and Exclusion Criteria \label{ssec:inclusion-exclusion}}

We formulated the inclusion and exclusion criteria in an iterative fashion during the initial meetings. Two authors applied the inclusion and
exclusions criteria in a two-phases strategy as suggested by Kitchenham \cite{Kitchenham:2007}. In the first phase, we inspected the paper titles. In the second phase, we inspected the abstracts and the conclusions.

\textit{Inclusion criteria}
\begin{enumerate}[noitemsep, nosep]
\item Answers our research and review questions
\item Suggests a software engineering method to improve one aspect of simulation software development
\item Is a research article  (including empirical studies,  review articles, opinion articles, and experience reports)
\item Published in the time range 1990-2015
\end{enumerate}

\textit{Exclusion criteria}
\begin{enumerate}[noitemsep, nosep]
\item Book
\item Database focused
\item Hardware focused 
\item Deals with simulation software for the improvement of software engineering
\item External to software engineering
\item External to simulation software
\item Education-related
\end{enumerate}

\subsection{Data Extraction and Tabulation}

All authors but one of the present paper worked on the data extraction and tabulation. The allocation of the papers was randomized and assigned equal workloads. We set biweekly meetings to share the status of the tasks and to reach agreements. We recorded the data in a Google Form, which resulted in a shared Google Sheet.

We extracted demographic data of each paper such as author 
names and their primary research areas, paper title, publication year,
and publication venue. We classified each paper as opinion,
practice/experience report, empirical with intervention (e.g.,
controlled experiments), or empirical without intervention. We extracted
the main research questions (or the research goals and objectives). 
We elicited the challenges addressed in the papers which were related to software engineering for simulations and the areas of software engineering as defined by the SWEBOK \cite{SWEBOK2014}.

Five questions synthesized the content of each paper. Two of those questions varied according to the article type. Three questions were in common.

\begin{itemize}[noitemsep, nosep]
\item \textit{Opinion or philosophical}---(1) What is the opinion of the author? (2) What is proposed?
\item \textit{Practice or experience report}---(1) What is reported? (2) What is the
potential benefit?
\item \textit{Empirical with intervention---}(1) What is the proposed solution? (2) What are expected advantages?
\item \textit{Empirical without intervention}---(1) What is the state-of-the-art? (2) What are the observations?
\item \textit{Common questions}---(3) What are the remaining problems, or did
nothing change? (4) Any potential remarks / anything else? (5) Which aspects of the SE/MP trade-off are covered and how?
\end{itemize}

\subsection{Quality Evaluation \label{ssec:quality-evaluation}}

Dyb{\aa} and Dings{\o}yr \cite{Dyba:2008fg}  defined eleven criteria for the quality evaluation of material included in SLRs. After a careful discussion, we agreed to include the following eight quality criteria (QC):

\begin{enumerate}[noitemsep, nosep, label=QC{\arabic*}]
\item \label{qc:1} Is this a refereed paper?
\item \label{qc:2} Is there a clear statement of the aims of the research?
\item \label{qc:3} Is there an adequate description of the context of the work?
\item \label{qc:4} Was the research design appropriate to address the aims of the research?
\item \label{qc:5} Was the recruitment strategy appropriate to the aims of the research?
\item \label{qc:6} Was the data collected in a way that addressed the research issue?
\item \label{qc:7} Has the relationship between researcher and participants been considered adequately?
\item \label{qc:8} Is there a clear statement and proof of findings?
\end{enumerate}

We assessed the quality of our sample using a short scale with the answers ``yes'', ``no'', and ``cannot tell'' to each criterion.

\subsection{SE/MP Trade-off}\label{ssec:se-mp-trade-off}

We analyzed the content of the papers for retaining those that mentioned any of the aspects dealing with the SE/MP trade-off. We extracted the SE/MP trade-off information and put it in an extra column of our shared spreadsheet. We met twice in the course of two weeks for discussing the filtered papers, the extracted data, and for agreeing on the steps to undertake.

\section{Results}

The search query that we input in the academic search engines and digital libraries yielded 768 papers, which we reduced to 255 after applying the first round of inclusion and exclusion criteria. The manual inspection of the related publication venues resulted in 86 papers after the first round of filtering. After adding the 69 QGS set of papers we reached 410 articles (255+86+69). A search fur duplicates let us reduce the dataset to 399 papers.

By applying the second round of inclusion and exclusion criteria, our dataset was further reduced to 82 papers. Finally, we identified 33 papers that deal with the SE/MP trade-off. The included and excluded studies are also available online \cite{Pflueger:2016}.

Several papers reported opinions and experiences. Paper \cite{P1} was written by software engineering researchers summarizing software risks for scientific software including a discussion of maintainability and portability and how they are linked. Efficiency is discussed in detail. There is even a mention of a 
trade-off: ``To the extent that advances in performance are achieved through hardware, software portability becomes more important than software efficiency.''
Paper \cite{P9} recommends practices such as to simplify and to organize the code to improve its maintainability. Paper \cite{P18} mentions all four quality attributes and
also discusses the trade-off based on their experiences in SSE. Paper \cite{P19}  is an experience report on the High Performance Computing Systems program of the Defense Advanced Research Project Agency (DARPA). The aim is to create an empirical basis describing expertise gaps in all aspects of productivity. The main gaps observed were developing HPC code, debugging/testing, optimizing, scheduling and using math libraries. Paper \cite{P22} contains opinions on the role of software infrastructure (compilers, languages, libraries or debuggers) for the overall productivity (including
 development effort, portability to next generation architectures, maintainability, performance, and scalability). Paper \cite{P32} discusses the risks of poor code documentation.

There were also four descriptions of surveys among computational scientists. Paper \cite{P6} describes a survey in which they identify efficiency as a main 
concern and state that maintainability is often poor. Paper \cite{P25} mentions maintainability: ``Software maintenance was generally ranked as moderately important.''
Paper \cite{P31} even claims that ``Code performance is not the driving force for developers or users; the science and portability are of primary concern.''
Paper \cite{P33} describes a survey in which they found: ``What is regarded as a priority in terms of software engineering are development methods, that help to design for performance from the beginning and to find the right (not too simple, not too complex) software architecture on the first run.''

The following papers went more into detail and performed what could roughly be categorized as case studies and experiments. Paper \cite{P8} describes 
a series of case studies where they found that performance competes with other important goals and that the low use of higher-level languages is a problem for portability. Paper \cite{P10} investigates in a pilot study the use of design patterns in parallel programming to increase maintainability. Paper \cite{P11} describes case studies which touch on several attributes and also mentions a trade-off while tuning the code between efficiency and maintainability. Paper \cite{P14} is 
about making legacy code parallel. Paper \cite{P15}  is a long-term analysis of a scientific software with a focus on its maintainability. Paper \cite{P16} deals with maintainability and performance. They wrap existing legacy-codes in architecture-aware interfaces to embed them into architectured software systems. 
This yields modular software systems maintaining the performance of the legacy code. Paper \cite{P20} contains a case study for using test-driven development
for cardiac simulation software. They focus on modularity and use of state-of-the-art numerics, and they develop a set of maintainable libraries.
Paper \cite{P23} is an analysis of commit comments in scientific software projects. Good commit comments help in maintaining software.

We found three articles proposing specific methods in the design of a software. Paper \cite{P21} proposes a tool for the automatic extraction of a 
UML class diagram from an object-oriented FORTRAN code. Paper \cite{P26} proposes and shows how to define good interfaces for scientific
software: ``We talk about how those interfaces allow us to generate most code automatically as an expert would manually.'' In paper \cite{P29}, the
authors argue that using test-driven development, the library becomes easier to use and extend.

Most articles propose or somehow describe the usage of particular frameworks, libraries or languages. Paper \cite{P2} mentions all four SE/MP attributes 
and discusses the proposed framework according to them. No trade-offs are mentioned. Paper \cite{P3} describes a case of using the OpenACC programming environment to improve maintainability, portability and performance. The OpenACC language is used to allow for portability while keeping efficiency. 
Paper \cite{P4} is similar to \cite{P3} and describes a smaller rewrite using \mbox{OpenACC}. Paper \cite{P5} proposes model-driven development for HPC and claims to support efficiency and portability. Paper \cite{P7} investigates how a DSL can provide better maintainability while keeping good efficiency. Paper \cite{P12} describes 
design patterns for FORTRAN for modern, maintainable software which is still efficient. Paper \cite{P13} reports on the experience of using existing 
frameworks especially for infrastructure and data management to improve extensibility while keeping high efficiency.
Paper \cite{P17} reports on experiences from the NSF FLAME project (on dense linear algebra library development). The approach is ``design by transformation'' 
using, e.g., domain specific languages to automate optimization based on codified experts' knowledge. Paper \cite{P24} talks about ``portable message-passing programming'' and parallelization and load balancing. Paper \cite{P27} shows the trade-off between efficiency and maintainability. It is a solution to a specific 
part of the trade-off: Choosing the best al\-go\-rithms/im\-ple\-men\-ta\-tions for a problem.
Paper \cite{P28} addresses performance and somehow maintainability: ``Dynamic scripting programming languages, in general, have distinct advantages in 
terms of developer productivity over compiled languages such as C/C++ and FORTRAN, because of their ease of use and extensive libraries. Such ``static'' languages have clear performance advantages due to their optimizing compilers.''
Paper \cite{P30} discuss the efficiency-portability trade-off and how to reduce it by libraries: ``The extensive usage of scientific libraries and parallel 
computing technology makes the programming easier and the model's computing performance much better. However this usage generally reduces 
software portability, which is one of the common problems for any large-scale, parallel scientific software.''

\begin{figure}[htbp]
\begin{center}
\includegraphics[width=.7\columnwidth]{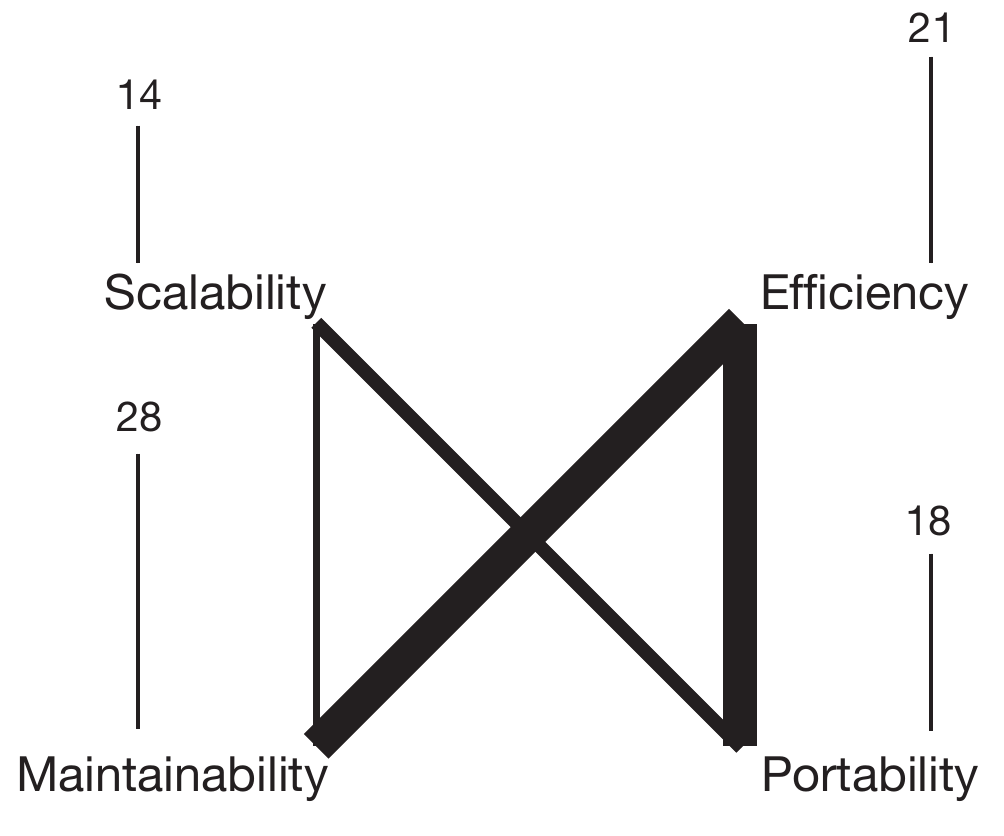}
\caption{An overview of how often the quality attributes and trade-offs are mentioned}
\label{fig:quality-attributes}
\end{center}
\end{figure}

Overall, we found that there is a considerable amount of papers that consider at least one of the quality attributes
we focus on. Fig.~\ref{fig:quality-attributes} shows that in 28 of the papers, maintainability played some role. Efficiency
was mentioned in 21, portability in 18 and scalability in 14 papers. The figure also visualizes the discussed trade-offs
between quality attributes. Trade-offs between efficiency and maintainability as well as between portability and efficiency
have been mentioned 10 times each. The trade-off between scalability and portability was mentioned in four papers, the trade-off
between scalability and portability only twice. There was, as expected, no trade-off between scalability and efficiency or maintainability
and portability.

\begin{figure}[htbp]
\begin{center}
\includegraphics[width=1\columnwidth]{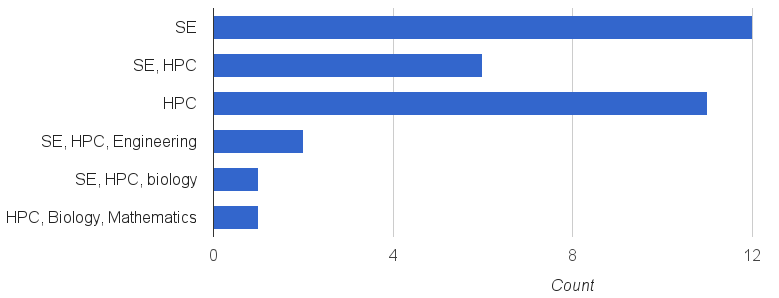}
\caption{The backgrounds of the authors in the included articles (SE: software engineering, HPC: high-performance computing)}
\label{fig:author-background}
\end{center}
\end{figure}

Most authors of our included articles have either a background in software engineering or high-performance
computing. Actually, as Fig.~\ref{fig:author-background} shows, most of the papers were authored by software engineering researchers 
only. The second most frequent group of papers was authored only by researchers from the high-performance computing domain. Yet, there are also
six papers with authors from both communities. The collaboration, however, could be significantly improved. Even worse, from the
communities the scientific software is designed for, there are only few authors: in our SLR, software engineering and high-performance
computing researchers collaborated with scientists from engineering, biology and mathematics.

Furthermore, we classified the publication venues of articles included. Tab.~\ref{tab:article-venues} shows the results. Almost half of the included
articles were published in workshop proceedings. The regular workshops on software engineering in science and engineering co-located
with the International Conference on Software Engineering (ICSE) and the International Conference for High Performance Computing,
Networking, Storage and Analysis (SC) are the venues for all of those. Then there is roughly the same number of journal and conference
proceedings articles, which is an expected distribution in computer science. Only one paper was published as a technical report and, hence,
definitely non-refereed.

\begin{table}[htb]
\caption{Article types of included papers, occurence, and references}
\begin{center}
\begin{tabularx}{\columnwidth}{lrX}
\toprule
\makecell[l]{Practice/experience \\ report} & 13 & \cite{P2}, \cite{P3}, \cite{P4}, \cite{P5}, \cite{P7}, \cite{P16}, \cite{P17}, \cite{P21}, \cite{P24}, \cite{P26}, \cite{P28}, \cite{P29}, \cite{P30} \\ \hline
\makecell[l]{Empirical study \\ without intervention} & 11 & \cite{P6}, \cite{P8}, \cite{P11}, \cite{P15}, \cite{P19}, \cite{P23}, \cite{P25}, \cite{P27}, \cite{P31}, \cite{P32}, \cite{P33} \\ \hline
\makecell[l]{Empirical study \\ with intervention} & 5 & \cite{P10}, \cite{P12}, \cite{P13}, \cite{P14}, \cite{P20} \\ \hline
\makecell[l]{Opinion/philosophical \\ paper} & 4 & \cite{P1}, \cite{P9}, \cite{P18}, \cite{P22} \\
\bottomrule
\end{tabularx}
\end{center}
\label{tab:article-types}
\end{table}%

We also classified the included article in different types. We show the results in Tab.~\ref{tab:article-types}. Most of the included
papers are practice or experience reports of some sort. They often describe some initiative or project from the scientific computing
domain. In the included papers, however, we can also find several empirical studies. Most of them without intervention; those are
usually surveys among scientists who develop software. Empirical studies with intervention are a kind of case study that applies
a new method, tool, framework or language. Finally, we also included four opinion papers that somehow argued about one of the attributes or
about a trade-off.

We report on our quality evaluation in Tab. \ref{tab:quality_evaluation_results}. At this stage, we focus on the counts of items for assessing each quality criterion and not their score. The reason is that more than half of our sample of SE/MP trade-off papers are of philosophical or reporting nature (see Tab. \ref{tab:article-types}). The scores for \ref{qc:4}, \ref{qc:5}, \ref{qc:6}, and \ref{qc:7} would be unbalanced by the non-empirical studies, or be a too small sample for meaningful interpretations if we ignore the non-empirical studies.

\begin{table}[htb]
\caption{Quality evaluation results (see section \ref{ssec:quality-evaluation})}
\begin{center}
\begin{tabular}{lrrrr}
\toprule
\textbf{Quality criteria} & \textbf{Yes} & \textbf{No} & \textbf{Cannot tell} \\
\ref{qc:1} & 24 & 3 & 6 \\
\ref{qc:2} & 31 & 0 & 2 \\
\ref{qc:3} & 26 & 5 & 2 \\
\ref{qc:4} & 16 & 1 & 16 \\
\ref{qc:5} & 8 & 4 & 21 \\
\ref{qc:6} & 12 & 5 & 16 \\
\ref{qc:7} & 4 & 8 & 21 \\
\ref{qc:8} & 16 & 10 & 7 \\
\bottomrule
\end{tabular}
\end{center}
\label{tab:quality_evaluation_results}
\end{table}%

\begin{table}[htb]
\caption{Publication venues of included papers}
\begin{center}
\begin{tabular}{lr}
\toprule
Articles in workshop proceedings & 16\\
Journal articles & 8 \\
Articles in conference proceedings & 6\\
Magazine articles & 2\\
Technical report & 1\\
\bottomrule
\end{tabular}
\end{center}
\label{tab:article-venues}
\end{table}%

\section{Discussion}

After summarizing the results of the SLR in the previous section, we discuss the findings and conclusions for future 
research that we draw.

\subsection{Principal Findings}

Our main finding is that \emph{we are not alone with the observation that the SE/MP trade-off plays an important role in
the development and maintenance of scientific simulation software}. Even in this initial SLR, we found
many articles mentioning it. Our review was not strictly restricted to simulation software but regarded the wider field of high performance applications.
Only very few contributions have a clear focus on simulation software with its 
specific requirements. However, in this wider field, \emph{the individual quality attributes are clearly recognized as important}.

Furthermore, we identified four different classes of contributions in the included articles as shown in Table \ref{tab:contribution-classes}.

\begin{table}[htb]
\caption{Classes of contributions in included articles, occurrence, and references}
\begin{center}
\begin{tabularx}{\columnwidth}{lrX}
\toprule
\makecell[l]{Describe/analyze \\ single attributes} & 7 &  \cite{P6},  \cite{P15},  \cite{P22},  \cite{P23},  \cite{P25},  \cite{P32},  \cite{P33} \\ \hline
\makecell[l]{Describe/analyze \\ trade-off} & 5 &  \cite{P1},  \cite{P8},  \cite{P11},  \cite{P18},  \cite{P31} \\ \hline
\makecell[l]{Solutions for \\ single attributes} & 8 &  \cite{P2},  \cite{P9},  \cite{P10},  \cite{P14},  \cite{P19},  \cite{P20},  \cite{P21},  \cite{P29} \\ \hline
\makecell[l]{Solutions for \\ trade-off} & 13 &  \cite{P3},  \cite{P4},  \cite{P5},  \cite{P7},  \cite{P12},  \cite{P13},  \cite{P16},  \cite{P17},  \cite{P24},  \cite{P26},  \cite{P27},  \cite{P28},  \cite{P30} \\
\bottomrule
\end{tabularx}
\end{center}
\label{tab:contribution-classes}
\end{table}%

The papers describing or analyzing single attributes concentrate on one or more of the quality attributes. 
We have seven such papers. 
They focus for example on the analysis of the maintainability of a scientific software over
time \cite{P15}. In contrast, papers that describe or analyze the trade-off are often based on surveys in which they found
something about the opinion of scientists on the trade-off (e.g.\ \cite{P33}). 
We have only five papers in this class, and none
gives any concrete means to analyze the effects of different decisions in the trade-off.

By far \emph{most of the papers present some kind of solution or solution proposal}. Eight articles provide solutions for one or more of the
quality attributes individually. For example, \cite{P9} describes some general guidelines to improve code to make it
more maintainable. Most solution proposals, however, try to actually improve the trade-off more or less explicitly. We have
13 papers in this class. For example, \cite{P2} and \cite{P3} report on OpenACC and its usage to improve existing simulation
code. Frameworks, libraries or specific languages are proposed to raise the level of abstraction (for easier maintenance
and faster or even automatic porting) while still being able to reach high levels of efficiency and scalability. In some cases,
performance measurements to support this claim are presented. Most evidence is anecdotal, however.

\subsection{Strengths and Weaknesses}

Overall, \emph{there is evidence for the SE/MP trade-off}. Half of the publications, however, result from workshops which often
accept less mature findings (including the present paper). 
Hence, for a part of these papers, we have to be careful about the validity. 
Also many of the papers have no clear empirical design but are opinion pieces or experience reports. A strong empirical foundation is yet missing. The topical conclusion is
that there is a lot more work to be done in the community to establish quantitative 
metrics and systematic methods to address the SE/MP trade-off. 

Furthermore, often we find only SE or only HPC researchers among the authors. This narrows the view of each individual
paper to the community the authors come from. This might be reflected in the findings.

\subsection{Meaning of Findings}

We see three main outcomes of these findings: (1) There has to be more research on the basic factors in the trade-off as 
well as its analysis/evaluation. The trade-off is mentioned often. Yet, \emph{we have no single quantification of the trade-off in a
concrete setting}. To be able to optimize it, we need to better understand it. This includes also aspects such as the process
or guidelines to follow to reach a good trade-off. There is nothing on these aspects in the included papers. Furthermore,
they often only report single or few cases in specific settings. \emph{It is yet unclear whether their insights are transferable to other
settings or generalisable to broader domains}.

(2) 
\emph{There is still a need for further research on practices, methods and tools 
to help reduce the trade-off} (keep scalability/efficiency while improving maintainability/portability). Especially in combination with
(1), we should develop those practices, methods and tools more targeted and supported by more quantitative metrics. 

(3) Finally, we see \emph{room for much more collaboration across scientific communities}. By bringing together the viewpoints and
experiences of not only SE and HPC researchers but also the scientists from mathematics, engineering, biology or physics who often develop simulation software, we can reach better and more comprehensive insights.

\section{Conclusion}

The emerging field of Simulation Software Engineering (SSE) fills a central gap in the development of (scientific) simulation software. We have motivated our research on the Scalability-Efficiency/Maintainability-Portability (SE/MP) trade-off, one of the core challenges of SSE, reporting on our own experiences. In a systematic literature review, we have examined to what extent this trade-off has been reflected in the literature. While there is evidence for the SE/MP trade-off, we conclude that significantly more effort is necessary in the research community: Anecdotal findings and solutions that are custom-tailored to specific problems have to be generalized; there is need for new metrics, practices, methods and tools; and there are too few collaborations across the scientific communities that are involved.

\section*{Acknowledgments}
One of the authors acknowledges support of the DFG within the Cluster of Excellence in
Simulation Technology (EXC 310/2).

\bibliographystyle{IEEEtran}
\bibliography{IEEEabrv,sse}

\end{document}